# Computational Content Analysis of Negative Tweets for Obesity, Diet, Diabetes, and Exercise

George Shaw, Jr., Amir Karami
School of Library and Information Science, University of South Carolina
1501 Greene Street, Columbia, SC 29208
gshaw@email.sc.edu, karami@mailbox.sc.edu

**ABSTRACT**
Social media based digital epidemiology has the potential to support faster response and deeper understanding of public health related threats. This study proposes a new framework to analyze unstructured health related textual data via Twitter users' post (tweets) to characterize the negative health sentiments and non-health related concerns in relations to the corpus of negative sentiments; regarding Diet Diabetes Exercise, and Obesity (DDEO). Through the collection of 6 million Tweets for one month, this study identified the prominent topics of users as it relates to the negative sentiments. Our proposed framework uses two text mining methods, sentiment analysis and topic modeling, to discover negative topics. The negative sentiments of Twitter users support the literature narratives and the many morbidity issues that are associated with DDEO and the linkage between obesity and diabetes. The framework offers a potential method to understand the publics' opinions and sentiments regarding DDEO. More importantly, this research provides new opportunities for computational social scientists, medical experts, and public health professionals to collectively address DDEO-related issues.

**INTRODUCTION**
Social media has provided an additional source that enables us to view, like, and disagree with opinions and experiences of those we know and those who we may not know. Social media also provides a platform to communicate information, market products, and for everyday individuals to express their sentiments (feelings) regarding multiple political, entertainment, and societal issues (Mejova, Weber, & Macy, 2015; Pang & Lee, 2008).

The ability to analyze various sentiments of social media users, provides candid insight to users' feelings and opinions. Social-media based digital epidemiology can support faster response and deeper understanding of public-health threats than traditional methods (Eichstaedt et al., 2015). The publicly available data on Twitter has created a new intersection of public health, health informatics, and data science. Automatic systems can probabilistically infer what is happening around the world by using the data of what people are thinking and doing (Nasukawa & Yi, 2003; Wiebe et al., 2003; Zabin & Jefferies, 2008). With roughly 328[1] million active users, Twitter is seen as a reliable data source to provide real-time feedback and opportunity to understand users' concerns (Mejova, Weber, & Macy, 2015).

While there are many social phenomena that we are able to determine by users' sentiments, common health issues will be the focus of our conversation. From 1980 to 2014, obesity has doubled according to the World Health Organization (WHO). Moreover, roughly 29% of the world's population is considered overweight and over 600 million adults are considered obese. The lack of physical activity, coupled with the excess intake of energy from food, contributes to weight gain and diabetes (Hill, Wyatt, & Peters, 2012; Wing et al., 2001). While we know that some populations are more affected than others, fears have risen in that this generation may have a shorter life expectancy than their parents (Flegal, Carroll, Kit, & Ogden, 2012; Ogden, Carroll, Kit, & Flegal, 2012; Wang, Beydoun, Liang, Caballero & Kumanyika, 2008).

A change in lifestyle behaviors, such as proper diet and exercising, can drastically alleviate the obesity public health epidemic through these modifiable behaviors (Wing et al., 2001). The lack of exercising and it's linked to obesity, increases the risk for heart disease, hypertension, diabetes, arthritis, and certain cancers (Blackwell, Bass, Bishop, & Hussaini, 2012). It is estimated that by 2050, roughly 29 million Americans will be diagnosed with Type 2 diabetes; which is a substantial increase of 165% from the 11 million diagnosed with the condition in 2000. Furthermore, several studies have shown the relations that exist among diet, diabetes, exercise, and obesity (DDEO) (American Diabetes Association, 2004; Barnard et al., 2009; Hartz, Rupley, Kalkhoff, & Rimm, 1983).

---

[1] https://www.statista.com/statistics/282087/number-of-monthly-active-twitter-users/

"Sentiment and subjectivity are quite context-sensitive, and, at a coarser granularity, quite domain dependent" (Pang & Lee, 2008, p. 21). Meaning, sentiment opinions are different across various domains and its intended use. Two exact expressions posted on two different webpages or social media platforms can elicit different meanings. Remaining cognizant of this issue when doing sentiment analysis research; sharing millions of posts in social media, such as Twitter, still provides a great opportunity to track people's health concerns.

This research proposes a framework, using two text-mining methods, to analyze health related text data via Twitter users' post (tweets). This study seeks to characterize the negative health sentiments and non-health related concerns in relations to corpora of negative sentiments regarding DDEO. Diabetes is a significant medical condition that is normally associated with overweight and obesity (Flegal, Carroll, Kit, & Ogden, 2012). Two behavioral factors that can reduce an individual's risk of being obese and consequently, developing diabetes, are proper dieting and getting the recommend daily physical activity suggested by the evidence-based Physical Activity Guidelines of America (Wing et al., 2001)[2]. Based on these studies, diet, diabetes, and exercise; in addition to obesity were chosen as topics.

While the focus of the framework will be applied to unstructured Twitter data, there is potential for this framework to be used on reports and notes generated from health information systems or web-based weight management interventions with social media components. The results from this framework can also offer health professionals' additional insight to individuals' negative sentiments about exercising regularly, dieting, and diabetic concerns. This study attempts to answer: *"What are the negative characteristics of health and non-health concerns of people with respect to DDEO?"*

**LITERATURE REVIEW**
Surveys and behavioral risk factor surveillance systems such as those utilized by the CDC are commonly used methods to collect health data. However, it can take weeks or years to collect, clean, and analyze the data (Wartell, 2015). Given the dynamic nature of Twitter, it offers researchers a new method to collect relatively real-time data (Eichstaedt et al., 2015). By using data from Twitter, we are able to identify sentiments and topics using sentiment analysis and topic modeling. The automatically identified topics and corresponding opinions provide a fine-grained understanding of opinionated text data that is done through social media to identify public opinion (Rahman & Wang, 2016; Jo & Oh, 2011; Lin & He, 2009, Paul & Dredze, 2012). This study defines negative sentiments as those opinions that express a desirable state away from a neutral positioning (Yi, Nasukawa, Bunescu, & Niblack, 2003; Nasukawa & Yi, 2003).

Researchers have shown that Social Networking Site (SNS) users, particularly Twitter, are more inclined to share their health information within Twitter than they would with their doctor (Paul & Dredze, 2014). SNS such as Facebook, Twitter, and Instagram have grown in their tracking of health issues and show strong correlation between the trends in social media and the U.S. Centers for Disease Control and Prevention reports (Paul & Dredze, 2011; Paul & Dredze, 2014). The following sections will describe how Twitter has been used to track and identify non-health and health-related issues and those studies that have specifically utilized Twitter for sentiment analysis and topic modeling.

**Twitter and Tracking Health Issues**
Twitter has been used to mirror the offline sentiments of the general public based on a content analysis; and its prediction power to determine the likely outcomes of the stock market based on mood analysis (Bollen, Mao, & Zeng, 2011; Tumasjan, Sprenger, Sandner, & Welpe, 2010). From a health perspective, Twitter has been used to address a number of issues. There have been numerous studies showing Twitter's ability to predict influenza outbreaks. In 2009, Twitter was instrumental in predicting the swine flu pandemic using n-grams historical context (Culotta, 2010; Lampos & Cristianini, 2012; Lampos, Bie, & Cristianini, 2010; Paul & Dredze, 2011).

A study by Abbar, Mejova, and Weber (2015) demonstrated that friends on Twitter normally live in close geographic context of each other and share similar demographic variables. Friends are also likely to show similar interest toward food. A similar study showed that people, who tweet about drug abuse, also have friends that are dealing with similar problems (Hanson, Cannon, Burton, & Giraud-Carrier, 2013).

Another study utilized Twitter to identify evidence regarding the influence of social media on public health and its use to communicate health information. Using #childhoodobesity to monitor private person, for profit, nonprofit, media, government, educational tweets, and network connections; the researchers found lack of government, media, and educational sources of childhood obesity (Harris, Moreland-Russell, Tabak, Ruhr, & Maier, 2014). This may suggest a limited presence of credible sources and information about childhood obesity on Twitter. In addition, the limited information regarding home or school based strategies for reducing childhood obesity show that limited information regarding these evidence-based strategies are available (Harris et al., 2014). These research studies demonstrate Twitter's ability as a tool to track, predict, and model offline behavior.

---
[2] https://health.gov/paguidelines/guidelines/summary.aspx

The analysis of Tweets can help allied health professionals identify users' negative sentiments as it relates to the characteristics of DDEO. This research identifies the prominent topics of users as it relates to the negative sentiments (feelings). Statements themselves may be negative and understanding prominent topics among negative sentiments (corpus of negative statements) is an alternative approach to identifying the topics related to DDEO based solely on negative expressions (Nasukawa & Yi, 2003).

**METHODOLOGY**
In conjunction with data collection, the proposed framework utilizes sentiment analysis and topic modeling (Figure 1) to identify the prominent negative expression of Twitter users by applying the topic modeling on negative sentiments.

The question of "What are the negative characteristics of health and non-health concerns of people with respect to DDEO?" will be addressed through the collection of a large number of tweets using the Twitter API (Application Program Interface).[3]

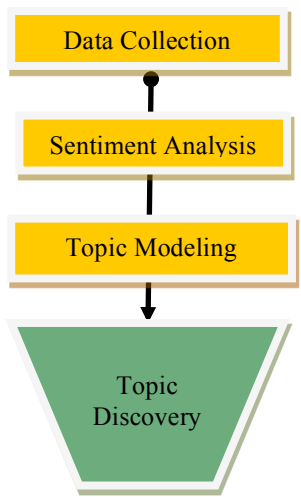

**Figure 1: Negative Analytics Framework**

**Data Collection**
This study collected the Twitter data using a real-time data collection method. By using this method, we are able to collect 10% of the publicly available English tweets. Previous research has analyzed a particular health topic such as influenza or mental health issues (Lampos et al., 2010; Eichstaedt et al., 2015). However, the data collected for this study will focus on the four areas key terms Diet, Diabetes, Exercise, and Obesity. From these key terms, the search terms for the Twitter API were created. Eight queries were used to collect nearly 6 million tweets with 61 million tokens between June 1, 2016 and June 30, 2016 (Table 1).

As part of the data collection process, post processing of the data collected requires cleaning by removing stop words such as *and*, *of*, *the* (based on a standard list of stop words). This will allow the topic modeling toolkit used for the discovery of topics, to correctly identify the topics for analytics purposes.

| DDEO TOPICS | TWITTER API QUERIES |
|---|---|
| DIABETES | #diabetes OR diabetes |
| DIET | #diet OR diet |
| EXERCISE | #exercise OR exercise |
| OBESITY | #obesity OR obesity |

**Table 1: Queries for Twitter API**

**Sentiment Analysis**
Sentiment analysis shows subjectivity and polarity in text data with two main approaches: learning-base and lexicon-base. Learning-based approach applies machine learning techniques to build classifiers from data. Lexicon-based approach uses a pre-defined dictionary of positive and negative words to find the frequency of positive and negative words. (Medhat, Hassan, & Korashy, 2014). In this research, we use Linguistic Inquiry and Word Count (LIWC) that is a Lexicon-based tool. This tool assisted with identifying the negative sentiments, based on the corpus of data collected. We found ~2 millions negative tweets.

**Discovery of Topics**
Topic discovery among the negative tweets to identify themes is done using topic modeling. As a commonly used method, the Latent Dirichlet Allocation model (LDA) was appropriate to use for this type of experiment. In addition to topic modeling abilities to cluster semantically related words, topic modeling has been used to discover relevant clinical concepts and structure in patient's health records, predict protein-protein relationships based on the literature knowledge, find patterns that exist within genetics data and image topic discovery (Arnold, El-Saden, Bui, & Taira, 2010; Blei, 2012).

LDA is based on the notion that the corpus contains topics and that each word in each document can be assigned to each of the topics with varying degree of weight per topic. In this experiment, we use LDA to identify the topics by using the corpus of negative tweets. Therefore, we utilize the results of the sentiment analysis for negative Tweets, employing topic modeling to discover the negative topics.

---
[3] https://dev.twitter.com/overview/documentation

We utilized the Mallet natural language processing program to identify topics in the negative tweets.[4]

After the data was cleaned, the model was trained to find the number of topics. To determine a proper number of topics, 80% of tweets for training and 20% of tweets for testing were used to estimate log-likelihood (Wallach, Murray, Salakhutdinov, & Mimno, 2009). Based on the log-likelihood estimation, we were able to determine that the optimum number of topics needed for training and testing was 425 topics (see Figure 2).

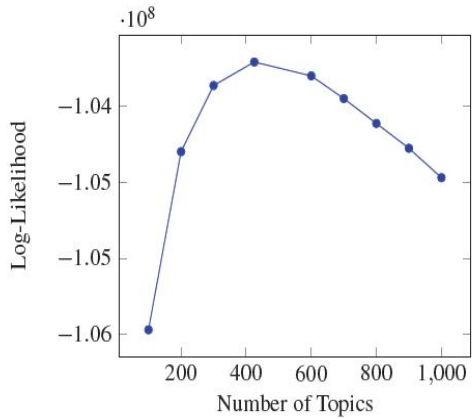

**Figure 2: Log-Likelihood Topic Estimation**

**RESULTS AND DISCUSSION**

There were unique sub-topics discovered in addition to topics that represented the keywords of our search. The discovered topics were assigned to one of the four search terms (DDEO words) that were used for data collection. The discovered topics that had more than half of the associated words relating to the DDEO words was labeled as Diet, Diabetes, Exercise, or Obesity. Table 2 shows the detected topics and sub-topics in the negative tweets. The main DDEO topic of exercise had a sub-topic of "mental health." We also see that the DDEO topics of exercise and diet are connected (Table 2).

**Obesity**

"Diet, diabetes, medical, and cancer" were the four sub-topics identified from the collected Twitter data as it relates to obesity. Based on the sub-topics, there are negative feelings towards things such as *bodyweight*[5] and obtaining *weight loss tips*. There was great concern regarding the impacts of obesity with relation to diabetes. Characteristics of "diabetes" included pre-diabetic (*prediabetes*), *arthritis*, and the *stress* with managing *weight* with foods high in *glycemic* (See Figure 3). Survey data shows that when individuals do not meet their individualized targets for glycemic control, blood pressure, or lipid control, they are at greater risk of diabetes (Ali et al., 2013).

| DDEO Topics | Sub-Topics |
|---|---|
| Diet | Food; Fastfood; Medications; Wellness; Alternative Diets; Religious Diets; Sweets |
| Diabetes | **Obesity**; Hypertension; Kidney; Cancer; Food; |
| Exercise | Lifestyle; Weightless; Body Image; **Diet**; Mental Health |
| Obesity | **Diet**; **Diabetes**; Medical; Cancer; Weight; Food |
| Non-Health Topics | People; Emotions; Celebrity; Government; Events |

**Table 2: Topics and Sub-Topics**

[6]

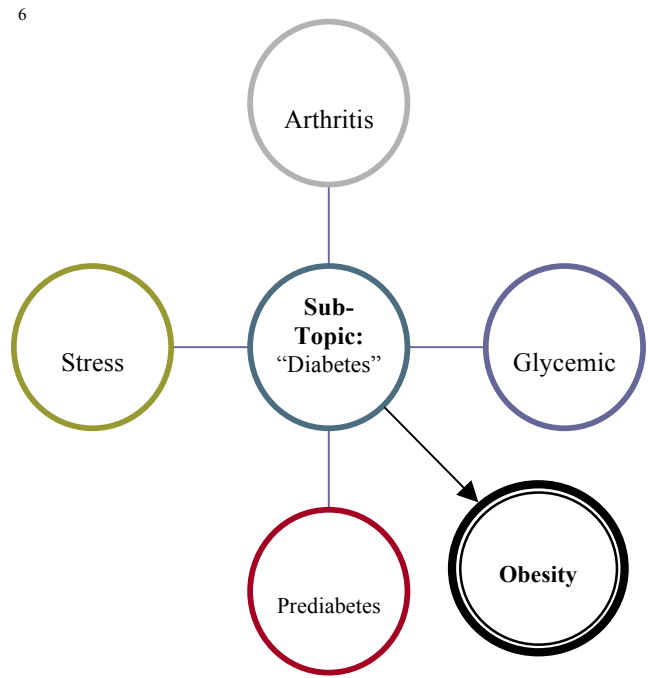

**Figure 3: Characterizing Words for Sub-topic "Diabetes" in relation to Obesity**

---

[4] http://mallet.cs.umass.edu/topics.php

[5] To assist readers with distinguishing between sub-topics and words that characterize the sub-topics, Quotation Marks "" are used to identify sub-topics words and *Italics* style is used to represent the words that characterize sub-topics

[6] Additional characterizing words included *lipolysis*, *weight*, and *diabetesawarness* (diabetes awareness).

There are also negative sentiments towards medical fads or conditions that promise *extreme weight loss*. *Gout* and pancreatitis (*paynecreatitis*) were listed as negative medical impacts of being obese. While there was no mentioning of specific foods, the inclusion of "diet" as a sub-topic does bring awareness to the nutritional impacts of food and obese individuals.

**Diabetes**
The most significant sub-topics identified included "obesity, hypertension, and kidney." Researchers have documented the link between obesity and diabetes (Harris et al.,2014; Harvard, n.d; Reilly & Kelly, 2011). The data reveal commonly seen results with characterizing words such as *excess fat*, difficulty *working out*, *belly fat*, and consuming excessive amounts of *sugar*. A surprising negative feeling about obesity was *genetic*. It is inconclusive, based on the sub-topics, the concern of genetics regarding obesity; but this may suggest a conversation involving the obesity pandemic that needs to be discussed. "Hypertension" (see Figure 4) was another sub-topic that yielded interesting characteristics. *Insulin* and *glucose* were concerns with regards to hypertension 2015). Continued issues with weight and high glucose levels can lead to kidney (renal) failure.

**Exercise**
Much of the public understands the connection between obesity and the health risks associated with the condition; such as diabetes and heart disease. Increased TV and computer time with lack of exercise are factors that contribute to obesity (Tompson et al., 2012). "Lifestyle" was one of the significant sub-topics identified with exercise. *Biking,* sporting activities (*sports*), and *running (*see Figure 5) were characterized as negative sentiments for engaging in physical activity. However, *soreness* and *inflammation* were also identified as words that characterized a lifestyle of physical activity.

These negative sentiments (i.e. *Biking, sporting activities*, and *running)* support data from the 2006 National Health Interview Survey in that "many overweight and obese people with prediabetes and diabetes do not report behaviors conducive to weight loss or physician advice for such behaviors" (Dorsey & Songer, 2011). The inability to change lifestyles may not only be a result of not engaging in physical activity, but dealing with the inflammation and physical aches (*soreness* and *inflammation* ) that are a result of being physically active and creates barriers to a physical lifestyle.

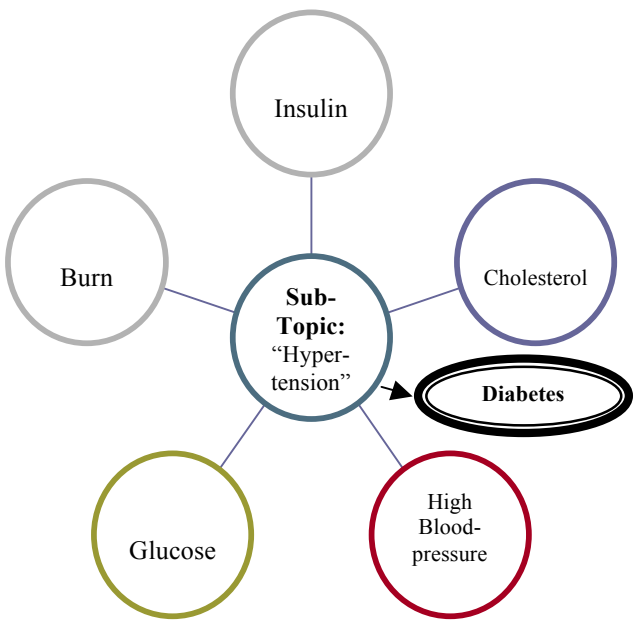

**Figure 4: Characterizing Words for Sub-topic "Hypertension" in relation to Diabetes**

A study in the 1980s, using random population sampling, identified the possible link between hypertension, obese individuals, and glucose intolerance (Modan et al., 1985). Furthermore, recent studies continue to show the link between diabetes and hypertension that affirms the negative sentiments Twitter users have regarding glucose levels and insulin (DeVallance, 2015; Lukic et al., 2014; Willig et al., "Mental Health" was another significant sub-topic identified with the exercise topic. Engaging in physical activity releases endorphins that trigger a positive feeling in the body, contributing to positive psychological benefits (Mead et al., 2009)[7]. However, being included as a negative sentiment, what do characterizing words such as *brain*, *alzheimer*, and *endorphins* imply for the sub-topic "mental health?" It is possible that individuals are concerned about their mental well-being due to the lack of exercising.

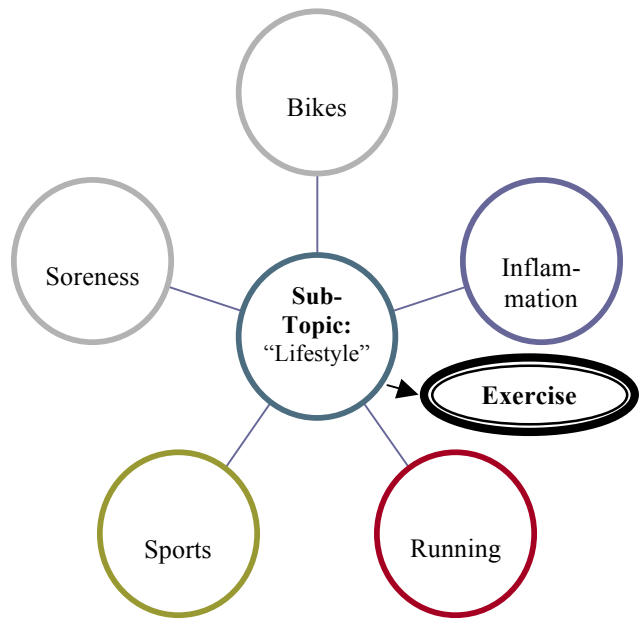

**Figure 5: Characterizing Words for Sub-topic "Lifestyle" in relation to Exercise**

**Diet**
Of the four main topics, Diet yielded the most sub-topics. Expected sub-topics based on previous studies included "food" and "wellness" (Abbar et al., 2015; Turner-McGrievy & Beets, 2015). Food tweeted by users are predictive of the national obesity and diabetes statistics (Abbar et al., 2015).

"Medications" was another sub-topic of diet that was intriguing. Considering that many fad diets advertise the use of some super pill to assist with losing weight within Twitter, a negative sentiment towards "medication" within this context is not alarming. However, *cancer* was one of the words that characterized medications. "Types of diets" was another significant sub-topic that described Diet. As seen in Figure 6, *nutrisystem*, *skinny diets*, and *vegan* are words that describe this sub-topic of diet. *Juicing* is a growing alternative diet trend that has conflicting sentiments regarding its effectiveness as a healthier option to eating fruits and vegetables (MacVean, 2015).

**Non-health Topics**
Examples of the non-health related topics based on our corpus can be seen in Table 2. While the framework did not perform extremely well with detecting non-health related topics, there were some topics identified that are applicable to the DDEO related tweets. Although "Emotions" was not identified as a sub-topic of our DDEO words, some of the characterized words for "emotion" (*feeling*, *piss*, *disabled*, and *burden*) may help explain how individuals who are struggling with being psychically inactive or have diabetes feel. This, again, supports Dorsey and Songer's (2011) study and how those with prediabetes and diabetes report behaviors to their doctor or primary care physician, despite the advantages of weight loss that can be achieved through lifestyle modification (Wing et al., 2001). Understanding the affective components of obese individuals can be beneficial to addressing their dietetic, diabetic, and physical activity concerns and behaviors.

---

[7] http://www.webmd.com/depression/guide/exercise-depression#1

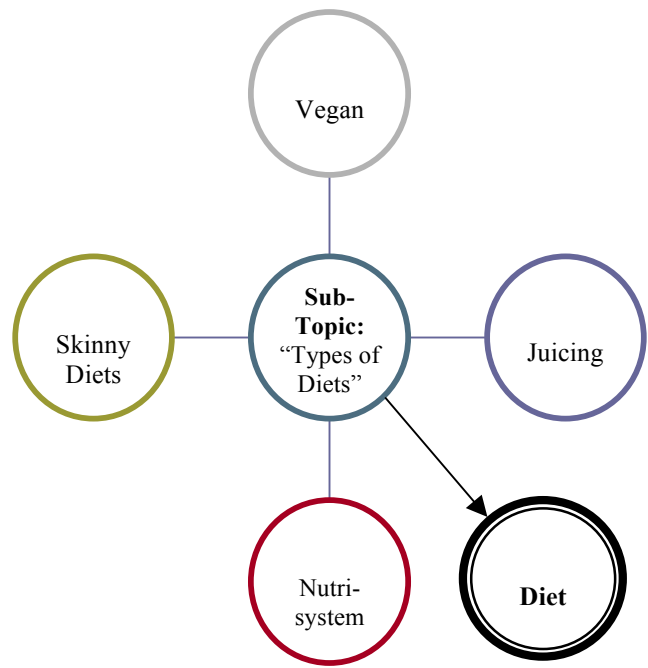

**Figure 6: Characterizing Words for Sub-topic "Types of Diets"**

**Analysis of DDEO Topics**
At least one DDEO related topic was represented for every topic, except Diet. There is a relationship that exists between the two topics, Diabetes and Obesity, as seen in Figure 7. Based on the characterizing words of the sub-topic "diabetes" for Obesity (ex. *prediabetes, weight glycemic, arthritis, stress*), this may suggest a negative opinion surrounding the conditions of being overweight or obese. Also, there appears to be a need for diabetes awareness (*diabetesawareness*).

Obesity included Diet as one of its sub-topics. With words such as *fatloss tips, veggie health tips*, and *fattening,* Twitter users are looking for help with their dieting efforts and weight control. Although the non-health related sub-topic of emotions was absent in Diabetes and Obesity, these emotions support the struggle and frustration that overweight and obese individuals deal with (Blackwell et al., 2012; Dorsey & Songer, 2011).

"Diet" was also a sub-topic of Exercise. The negative association of "body image" was a reason to *lose weight* and avoid being labeled a *fatty*, while dealing with the *frustration* and motivation to exercise (Dorsey & Songer, 2011). However, a caveat with this analysis is the notion of sarcasm and slang. As in many sentiment analysis studies, it is difficult to account for this complexity, as in some contextual settings, the use of the word fatty may suggest hunger and not indicative of an individual's physical or medical state, based on the meaning of the word[8].

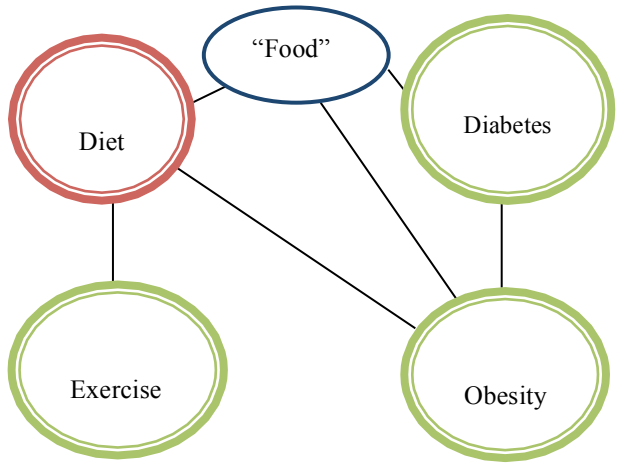

**Figure 7: Relationship among the DDEO Topics**

As previously noted, Diet did not contain any of the other DDEO topics, however, it contained a sub-topic ("food") that was also associated with Diabetes and Obesity. "Food" and "fastfood" were found in diet and the characterizing words of *fries*, *doritos*, *Mcdonalds*, *beer*, show foods high in sodium, sugar, or no nutritional value. "Nutritional status is an integral component of and has vital implications on the health of individuals," (Boumtje, Huang, Lee, & Lin, 2005, p. 116). Consistently eating the proper nutritional food is important with preventing excessive spikes in your glucose levels and preventing insulin intolerance (DeVallance, 2015; Lukic et al., 2014).

The relationship of "food" among the topics also illustrates a consistent factor with the literature regarding food, exercising and maintaining a healthy weight. Exercising is important, but those efforts are negated when our eating habits remain unchanged (Marks, 2004; Oaklander, 2015). More importantly, these results provide health educators with the covertly, yet blaring alarm of proper nutrition with regards to dieting efforts to prevent obesity that are not necessarily factors of genetics, psychological well-being, or environment.

Extensive analysis of the results from our framework can elicit numerous questions and analyses; such as the brief, aforementioned analysis demonstrated (coupled with supporting literature). These results also provided policy makers, health professionals/educators, and dietitians with reliable, significant probabilistic results that are based on millions of user provided information. Whether through direct topic or sub-topics and their characterizing words, in our case, our framework assisted with adding an intricate narrative to the negative content regarding obesity, and latent factors that help explain these negative sentiments.

**CONCLUSION**

The negative sentiments of Twitter users support the literature narrative and the many morbidity issues that are associated with obesity, diabetes, diet, and exercise. While diet was not topically associated with diabetes, exercise, and obesity, the sub-topic of "food" reveals the connection that it has with diabetes and obesity (Figure 4). Characterizing words such as *fries*, *beer*, *Doritos*, and *cookies*; are high in saturated fat, sugar, and excess calories (American Heart Association, 2017; Oaklander, 2015).

The negative sentiments in the study supports Arizona's 2012 Behavioral Risk Factor Surveillance System survey results and the Centers for Disease Control and Prevention statistics with regards to the negative impacts of not being physically active and the linkage between obesity and diabetes (Blackwell et al., 2012). This paper identifies the negative sentiments of Twitter and supports the concern that public health experts have regarding the obesity epidemic, based on literature and survey results. A majority of the public understands the connection between obesity, proper exercise, and the health impacts (Tomspon et al., 2012). In spite of the conversation and academic literature that has documented this issue, there seems to be an unclear direction with how to collectively address this problem among academic domains.

Two-thirds (67%) of adults' favor requiring chain restaurants to list calorie counts on menus, however, only 31% support limits on the size of sugary soft drinks in restaurants and convenience stores - 67% oppose this idea (Pew Research Center, 2013). "Government" was revealed as a negative non-health topic. While there is a clear understanding by the public to increase transparency and ingredients placed in foods, government restrictions on these issues are often faced with backlash.

Some form of a medical condition or concern was expressed in each of the DDEO related topics. This included characterizing words such as *cancer*, *diabetes*, *arthritis*, and *stress*. In many of these cases, the uses of drugs are prescribed to help individuals control or alleviate the medical condition. Most of the cost that is associated with treating obesity is related to the diseases that are associated with obesity and those required prescription drugs to treat the diseases or medical condition (Finkelstein, Trogdon, Cohen, & Dietz, 2009). This research study provides public health advocates with supporting evidence of the severity of obesity, particularly from a prescription drug standpoint. If this trend continues, 1 and every 6 dollars spent will be for healthcare reasons (Wang et al., 2008). Our approach to identify negative sentiments regarding obesity demonstrates the ability to discover common concerns associated with obesity and medical cost.

---

[8]characterized by overproduction or excessive accumulation of fat: http://www.dictionary.com/browse/fatty?s=t

Instead of using a linguistic only computational analysis that is commonly seen is this type of research, we were able to identify negative public health sentiments and opinions in DDEO related tweets using sentiment analysis. The sub-topics presented, represented the vastness of issues that concerns users with regards to DDEO. Characterizing words such *lipolyis* to acupuncture (*acupunture*), our framework was able to capture many sub-topics that would assist public health professionals. This research shows that computational scientists, medical experts, and public health professionals need to work on a united front to address issues regarding obesity.

This research has two specific limitations: time and space. The time limitation is that the data were collected during a one-month time period, June 1 – June 30 2016. We will need to extend the data collection period to address seasonal and temporal trends that impact our DDEO words used for collecting the data.

**ACKNOWLEDGMENTS**

We would like to thank the Information Technology staff (Jill Chappell-Fail and Jeff Salter) for the technical support and School of Library and Information Science Ph.D committee for supporting and mentoring the doctoral student that participated in this research study.